# A mid-infrared Mueller ellipsometer with pseudo-achromatic optical elements

E. Garcia-Caurel*, A. Lizana, G. Ndong, B. Al-Bugami, C. Bernon, E. Al-Qahtani, F. Rengnez, and A. de Martino († August, 2014)

LPICM, Ecole polytechnique, CNRS, 91128, Palaiseau, France.
*Corresponding author: enric.garcia-caurel@polytechnique.edu

The purpose of this article is to present a new broadband Mueller ellipsometer designed to work in the mid-infrared range, from 3 to 14 microns. The Mueller ellipsometer, which can be mounted in reflection or in transmission configuration, consists of a polarization state generator (PSG), a sample holder, and a polarization state analyzer (PSA). The PSG consists in one linear polarizer and a retarder sequentially rotated to generate a set of four optimal polarization states. The retarder consists in a bi-prism made of two identical Fresnel rhombs disposed symmetrically and joined by optical contact, giving the ensemble a "V" shape. Retardation is induced by the four total internal reflections that the beam undergoes when it propagates through the bi-prism. Total internal reflection allows to generate a quasi-achromatic retardation. The PSA is identical to the PSG, but with its optical elements mounted in reverse order. After a measurement run, the instrument yields a set of sixteen independent values, which is the minimum amount of data required to calculate the Mueller matrix of the sample. The design of the Mueller ellipsometer is based on the optimization of an objective criterion that allows minimizing the propagation of errors from raw data to the Mueller matrix of the sample. The pseudo-achromatic optical elements ensure a homogeneous quality of the measurements for all wavelengths. The performance of the Mueller ellipsometer in terms of precision, and accuracy, is discussed and illustrated with a few examples. © 2012 Optical Society of America

## 1. Introduction

Standard spectroscopic ellipsometry is an optical characterization technique sensitive to the polarization of light, which is also non-invasive and non-destructive. Standard ellipsometers have been used for more than one century to characterize the optical properties of materials and also to measure thickness of coatings made of single films or stacks. Former ellipsometers were limited to work in the visible spectral range, but decades of work in instrumentation together with the progressive availability of new optical components allowed to push the spectral limits to the vacuum ultraviolet and to the infrared [1]. There are two main reasons to explore the infrared range. The first one is the infrared absorption due to free carriers, it extends over a wide spectral range and depends on free carrier volume density and mobility. This feature can be used to assess the electrical properties of thin films made either of metals, semiconductors, or transparent conductors, which are commonly used to build low-emissivity windows, transparent electrodes for optoelectronic devices such as solar cells, light emitting diodes or liquid crystal displays. The second characteristic of the infrared is the occurrence of well-localized and spectrally narrow absorption bands due to the resonant excitation of vibrational modes of single chemical bonds in molecules or collective oscillations in networks of atoms. Those resonances are highly specific of the type of chemical bonds and allow to study the chemical composition of samples. The room for ellipsometry appears because the absorption due to chemical bonds is highly directional. Directional absorption creates linear dichroism that is easily measured in polarimetry. Linear dichroism can then be used to evaluate structural properties of samples such as the degree of molecular ordering or the degree of crystallinity. The ordering of molecules can also induce birefringence which is easily measured by polarimetry and can be used as an indicator to characterize sample properties. Nowadays, there are several relevant industrial and scientific activities involving the use of polymer thin films such as: pharmaceuticals, alimentary packing, organic light emitting diodes synthesis, or flexible panel display fabrication that can profit from polarization sensitive optical measurements in the infrared [2-4].

Despite of its great interest, standard ellipsometry, is limited to non-depolarizing samples, i.e. samples which do not introduce any randomness to the temporal or spatial vibration of the electromagnetic field associated to a radiation beam. Mueller ellipsometry, does not suffer from this limitation and can be used to study non-depolarizing as well as partially depolarizing samples. Mueller and standard ellipsometry provide equivalent results for non-depolarizing samples. Inhomogeneity in the samples is an important and rather common cause of depolarization. Inhomogeneity can appear either in sample chemical composition or structure (thickness, roughness or porosity). Because samples currently found either in everyday life or manufactured by the industry, may not be perfectly homogeneous, they may induce depolarization to a given extent. Therefore, the use of Mueller ellipsometry appears as an interesting alternative to extend the characterization capabilities of ellipsometry.

In the past ten years a number of spectroscopic Mueller ellipsometers have been developed. However, the vast majority of them were designed to work either in the visible [5-8] or in the visible-near infrared spectral range [9]. A few of them can make measurements in the infrared spectrum [10-15]. Our motivation was to develop a wide-spectral Mueller ellipsometer optimized to work in the mid-infrared to combine the sensitivity of ellipsometry to the physical structure of samples, with the chemical sensitivity characteristic of the infrared range. Such a Mueller ellipsometer will probably satisfy the growing demand of academic and industrial scientists and technicians, who need new information to characterize their samples. The instrument was patented a few years ago [16] and it was used in several studies [17-18]. The aim of this article is to present a detailed description of the instrument design and operation in a format different from that of a patent, understandable to those who are not familiar with legal documents. The characteristics that make this apparatus different to others, is the simple operation protocol, based in the non-continuous rotation of two retarders, and the use of achromatic optical elements to guaranty an optimal and uniform quality of measurement over the measured spectral range. In the first part of this paper we describe the design principle and the practical implementation of the Mueller ellipsometer. In the second part we show the results of some tests to assess the performances of the apparatus in terms of precision and accuracy. Finally, in the third part we provide two examples to illustrate representative results that can be obtained with the Mueller ellipsometer.

## A. Optical configuration and instrument operation

### 1. Optical configuration

The basic configuration of a Mueller ellipsometer consists of a light source, an input arm, a sample holder, an exit arm, and an acquisition system. In order to preserve the achromaticity of the system, gold coated mirrors have been used to define the optical path. The input arm includes an infrared source, a polarization state generator (PSG) and a retractable sample-holder for calibration samples. The exit arm includes also a mobile sample-holder for calibration samples, a polarization state analyser (PSA) and a detector system. The illumination source is a commercial FTIR spectrometer providing an infrared beam in a continuous spectral range from 2.5 to 14 μm. The PSG comprises a fixed grid type linear polarizer, and an achromatic retarder mounted in a rotatable holder. The rotatable holder allows orienting the retarder axis at four consecutive positions respect to the transmission direction of the fixed polarizer. Each one of the four orientations gives rise to a particular polarization state. The PSA is identical to the PSG, but with the optical elements, (retarder and polarizer) positioned in reverse order. Each one of the four polarization states created by the PSG, after being transformed by the sample, is projected to four optical configurations sequentially created by the PSA. Each analysis configuration is made by orienting the PSA retarder respect to the transmission axis of the polarizer at a given angle. Similarly to the PSG, the polarizer in the PSA is kept still during all the measurement process to avoid artifacts created by polarization sensitive detectors.

Although measurements can in principle be performed either in reflection or in transmission, we have chosen to work in reflection because it is the most sensitive to measure thin films on isotropic substrates. In Fig. 1 we have represented two possible variants of the Mueller ellipsometer. In one configuration the beam incident on the sample is collimated, whereas in the other configuration the beam is focused.

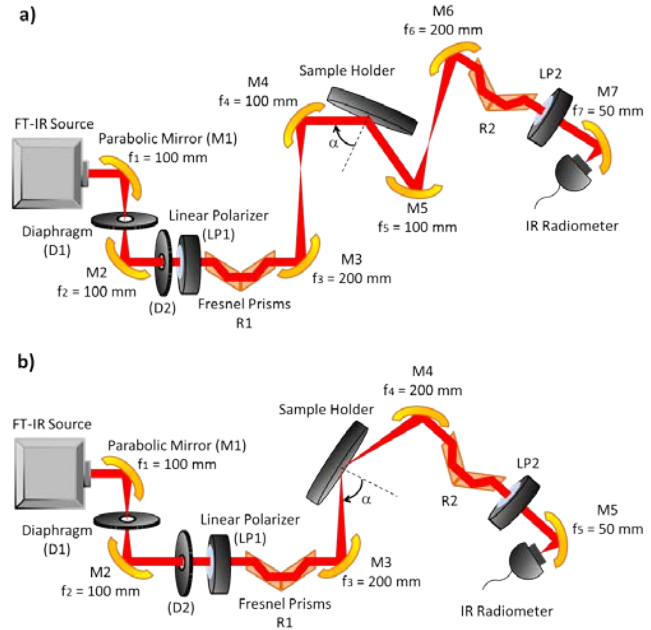

Fig. 1. Schematic representation of the optical configuration of the infrared Mueller ellipsometer: (a) collimated beam configuration and (b) focused beam configuration.

The collimated beam configuration is adequate to measure gratings or samples showing diffraction or scattering. This configuration provides a well-defined angle of incidence with controllable divergence (usually it is set to be around 1mrad) thanks to the aperture diaphragm (D2) placed at 100 mm from the second mirror (M2). The spot size on the sample, which is half of that of the aperture diaphragm because of the conjugation of focal lengths of mirrors M3 and M4, is 2x(2/cos(AoI)) mm. Here, AoI stands for angle of incidence. The spot size is not circular because of the oblique incidence. The focused beam configuration is well adapted to measure specular samples. It has the advantage of providing a smaller spot size than the collimated configuration, and the possibility to collect more light because the aperture diaphragm can be reasonably opened. The disadvantage of the focussed configuration is that the angle of incidence is less well-defined than in the collimated configuration. Indeed, focalization implies the existence of a distribution of angles of incidence across the focused beam waist, estimated to vary from 0.5 to 0.1 degrees depending on the diameter of the aperture diaphragm used. These values are related to the size of the beam and the focusing mirrors used in our set-up. Uncertainty related to the angle of incidence can be taken into account when modelling experimental data [1].

The angle of incidence is not fixed and can be varied. Our first attempt was to use the classical 'goniometric' method which consists of rotating the sample and the exit arm by an angle α and 2α respectively. However, we disregarded this method because it was very difficult to keep the instrument properly aligned. Our second attempt was inspired but not identical to the one discussed by Matsumoto et al. [19]. To change the angle of incidence by ω deg., the parabolic mirrors placed just before and after the sample must be pivoted by an angle ω and –ω respectively. After pivoting the mirrors, the sample must be shifted by a given distance Δz to compensate for the beam displacement. In practice, Δz is the distance required to keep the beam impinging on the same position of the sample surface, and to maximize the signal level measured by the detector. The method is sketched in Fig. 2. The main difference of our method and the one discussed in [19] is that we worked with a collimated beam instead of a focused beam. In our case, the advantage of using a collimated beam is that the variation of the distance, Δz, does not affect either the alignment nor the divergence of the beam across the exit arm of the Mueller ellipsometer.

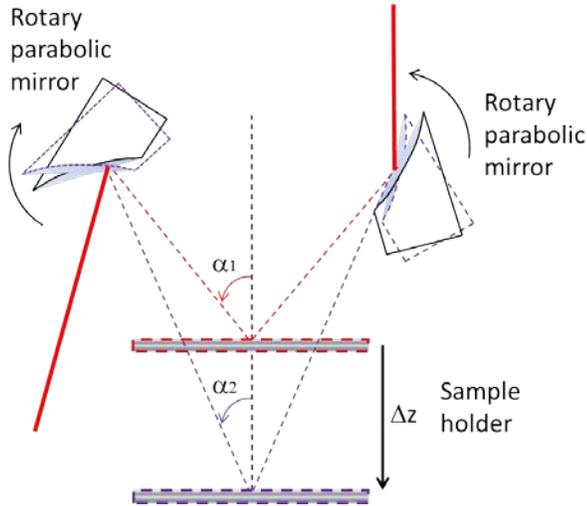

Fig. 2. Schematic representation of the different movements needed to modify the angle of incidence.

## 2. Instrument operation and optimization

As previously stated, the PSG generates a sequence of four independent polarization states. After reflection on the sample, the polarization states are projected on the PSA, which analyses with a sequence of four independent polarization configurations each polarization state arriving to it. As a result, a complete measurement yields a set of 16 independent intensity values that allow the calculation of the Mueller matrix. The four states of polarization generated by the PSG, as well of the optical configurations of the PSA, are not obvious or intuitive and they have been selected according to an objective criterion, called the maximization of the condition number [20], which allows the minimization of the propagation of errors from the measured raw-data to the computed Mueller matrix.

In the following we review the physical model used to design the optical setup of the PSG and the PSA and to define the condition number. It is common to use two 4x4 matrices, W, and A, to represent the PSG and the PSA respectively. The columns of the W matrix correspond to the four Stokes vectors generated by the PSG. Reciprocally, the analysis configurations used by the PSA are represented by four transposed Stokes vectors written as rows in the A matrix. The result of a polarimetric measurement consists of sixteen intensity measurements that can be expressed in terms of a matrix:

$$B = AMW, \quad (1)$$

where M is the Mueller matrix of the sample.

If A and W are known, then M can be extracted from the raw data B as:

$$M = A^{-1}BW^{-1}. \quad (2)$$

The determination of A and W is called calibration of the Mueller ellipsometer. Obviously, the matrices A and W must be non-singular. Moreover, in order to minimize error propagation from B to M, the analysis and modulation matrices, A and W respectively, must be "as close as possible" to unitary matrices. The best criterion to preserve A and W from singularity, is to optimize their condition numbers c(A) and c(W), which we define as the ratio of the smallest over the largest singular value of matrices A and W respectively.

In order to study the dependence of the condition number with the representative parameters of a given optical setup, a good approach is to create a functional model. Our PSG consists of a fixed linear polarizer followed by a retarder that can be rotated and placed at four different orientations respect to the polarizer. To simulate the optical behaviour of the PSG we treat the polarizer and the retarder as ideal components. The ideal Mueller matrix of the polarizer, P, and that of a retarder with a phase-shift "δ", and azimuth "θ", C(θ, δ), can be found in many publications and books such as "Handbook of Optics" [21] or [22] which focusses on polarimetric instrumentation. Assuming that the light beam entering the PSG is completely depolarized, the Stokes vector of the vector leaving the PSG is given by the expression:

$$S_{out}(\theta, \delta) = C(\theta, \delta) \cdot P \cdot (1 \quad 0 \quad 0 \quad 0)^T. \quad (3)$$

The W matrix can then be built by taking four Stokes vectors characterized, each one, by a given azimuth $\theta_i$ for i=1 to 4, and a common value for the retardation "δ". Under the above representation, the condition number of the matrix W can be understood as a function of five parameters, the four orientations and the retardance. As shown by J. S. Tyo [23], the theoretical value of the condition number for a PSG or a PSA consisting in one polarizer followed by at least one retarder, can vary between "0" and "$3^{-1/2}$". Accordingly, the values of the five parameters (retardation and azimuths), used to design the PSG, must be those that optimize the value of the condition number.

Finding the value of the parameters that maximize the condition number is a numerical problem that can be solved by means of standard numerical optimization packages. The resulting values for the retardation are: either 132°+n180° or 228°+n180°. The best sets of azimuths are: {$\theta_1$, $\theta_2$, $\theta_3$, $\theta_4$} = {38°+n180°, 74°+n180°, 106°+n180°, 142°+n180°}, "n" being a positive integer.

The tolerance of the absolute value to the retardation, as well as the four orientation angles of the retarder can be estimated with the help of the condition number. It is conventionally accepted that a minimum value of 0.25 is needed to obtain measurements with acceptable quality. Accordingly, the tolerances of the phase-shift, and the four azimuths of the retarder around their respective optimum value, are specified in order to keep the condition number higher than 0.25. The resulting tolerance range for the retardation and the orientation angles is ±15° and ±20° for respectively.

In view of the constraints imposed to the value of the retardation by the optimization of the condition number, and the wide spectral range intended to be measured, an achromatic retarder is needed. To achieve this objective, a total internal reflection (TIR) based achromatic retarder has been designed and built. The retarder consisted of a bi-prism made of two identical Fresnel rhombs disposed symmetrically and cemented together giving to the ensemble a "V" shape. An interesting review of alternative prism shapes has been written by J. M. Bennet [24]. The "V" shape bi-prism induces four total internal reflections and prevents the beam emerging from the bi-prism to be deviated from its original direction. The total retardance created by such "V" shaped bi-prism in contact with the air is given by:

$$\delta_{bi-prism} = 8\tan^{-1}\left(\frac{\cos\phi\left(n^2\sin^2\phi-1\right)^{1/2}}{n\sin^2\phi}\right), \quad (4)$$

where $\phi$ is the internal reflection angle of incidence, and "n" is the refraction index of the bi-prism. The prisms have been built in ZnSe because this material is transparent over a wide spectral range in the infrared. Because of the refractive index of ZnSe in the infrared, 2.429, the prisms must be cut to an angle of 58.25° to get the optimal retardation of 228°.

The strategy chosen to calibrate the broadband Mueller ellipsometer is based on the eigenvalue method (ECM) described by E. Compain et al. [25] because of its robustness. The ECM requires the measurement of a set of reference samples, called by some authors calibration samples. According to Compain's method we used the following calibration samples. A linear polarizer oriented respectively at 0° and 90° to the plane of incidence, and an achromatic retarder providing a total retardation of 110° oriented at 30° to the plane of incidence. We used as calibration retarder another "V" shaped prism. The calibration samples were placed between the sample holder and mirror M4 (see Fig. 1(a)) to calibrate the PSG, and between the sample holder and mirror M5 to calibrate the PSA (see Fig. 1(a)).

### B. Experimental validation of the optimized optical configuration

In this section, we show some results to discuss the conformity of the real Mueller ellipsometer to the theoretical model used to optimize the optical configuration. Moreover, based on the variance-covariance matrix formalism we provide a detailed evaluation of the noise (random error) propagation to each element of experimental Mueller matrices. This approach is useful to evaluate the uncertainty affecting each Mueller matrix element, in other words, the quality of measurements provided by the system. Fig. 3 shows the condition number of two typical experimental W and A matrices determined after calibration. As expected, the condition number of both, W and A, is quasi-achromatic. The slight spectral dependence observed in Fig. 3, can be attributed, according to Eq.(4), to the refractive index dispersion of the retarders. The spectral dependence is slightly higher for c(W) than c(A), but in practice those differences have a negligible impact on the quality of data. In our case, c(W) values vary between 0.40-0.45 and C(A) varies between 0.44 and 0.46, which are lower than the optimal $c_{max} = 3^{-1/2} = 0.58$ predicted by the ideal model and targeted in the design. This fact is explained because the real shape of the prisms does not correspond exactly to the optimal. Indeed, the prisms were cut to an angle of 60° instead of an ideal angle of 58.25° (Eq.(4)). Accordingly, the real retardance provided by both retarders was measured to be 215°. However, as it will be discussed in the following, even though the real condition number is lower than the ideal one, the Mueller ellipsometer still provides measurements of high quality.

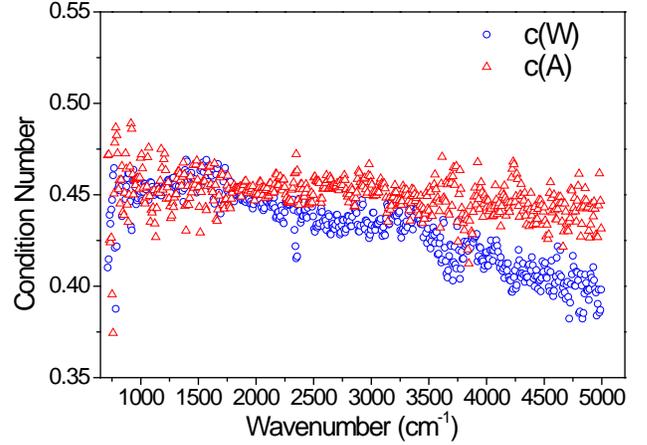

Fig. 3. Spectroscopic dependence of the condition number for experimental matrices W (blue dots) and A (red dots) in the range from 700 to 5000 cm$^{-1}$. Spectral resolution: 16 cm$^{-1}$. Number of averaged spectra: 16.

The condition number formalism gives an overall estimation of noise propagation from the intensity matrix, B, to the Mueller matrix, M. It fails, however, to provide a particular estimation of noise propagation for each Mueller matrix element. This drawback can be solved using an alternative approach based on the variance-covariance matrix shown in [26]. In the following, it will be assumed that the data of each matrix element in B, is characterized by a Gaussian noise with null mean and constant standard deviation $\sigma_0$. It is also assumed that the variance is independent of the absolute intensity measured by the detector, thus it is possible to assume the same value, $\sigma_0$, for the all the elements of matrix B. Accordingly the variance of each Mueller matrix element can be expressed by:

$$\forall i \in [1,16], \sigma_i^2 = \sigma_o^2 \left[\left[AA^T\right]^{-1}\right] \otimes \left[\left[WW^T\right]^{-1}\right]_{i,i}. \quad (5)$$

In Eq.(5) the matrices A and W, have been rewritten as 16 component vectors, according to the dactylographic convention. Noise propagation can be evaluated as the square root of Eq. 5. In Eq. 5 the factor following $\sigma_0$ determines by how much the variance characterizing the noise in matrix B will be amplified by the effect of matrices A and W. Therefore, we called the square root of this factor, 'noise propagation factor'. In the following example, we show an evaluation of the noise propagation factor from intensity matrices to Mueller matrices in three different cases represented by a Mueller ellipsometer working with: i) W and A matrices with an ideal optimal condition number equal to 0.58. ii) our experimental A and W matrices, and iii) W and A with a poor condition number equal to 0.25. In all three cases $\sigma_0$ is assumed to be equal to 1 for simplicity. Results are shown in Fig. 4. To evaluate noise propagation in practical situations it is necessary to use an estimation of $\sigma_0$, obtained from real data, instead of the value 1 used for this particular example.

As expected, the experimental noise propagation factors (case ii) are very close to the optimal ones (case i). The small differences among them are due to the fact that the condition number of the experimental A and W is not strictly optimal for the reasons discussed previously. Fig. 4 clearly shows that the accuracy obtained with our Mueller ellipsometer is close to the optimal one and definitely better than that obtained with a poorly conditioned system (case iii). Data in Fig. 4 also shows that even for the optimal configuration, the variance is not equally distributed for all Mueller matrix elements. Accordingly, noise propagation affecting the elements of the first column and row, is lower than that for the remaining matrix elements. This information may be of interest for instance in practical applications for which experimental data have to be fitted to a model. In the latter cases, being able to evaluate the uncertainty for each Mueller matrix element helps to decide whether the model accurately fits the experimental data or not.

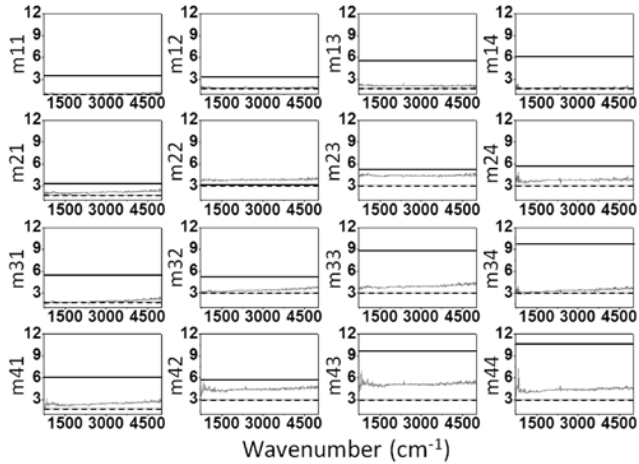

Fig. 4. Noise propagation factors for each Mueller matrix coefficient for the three situations described in the text: dashed line (optimum condition number), grey line (experimental matrices) and black line (poorly conditioned system). Scales are kept identical to highlight the differences among the three cases and also to make clear that noise propagation is not identical for all the elements.

As previously discussed, to quantify the accuracy of the Mueller ellipsometer, it is necessary to evaluate $\sigma_0$ from real data. This can be achieved easily by measuring a set of N independent intensity B matrices and then evaluating, for each matrix element, the corresponding standard deviation. Since we worked with spectroscopic data we had to repeat the evaluation for each spectral point. To study the dependence of $\sigma_0$ with N, we calculated $\sigma_0$ for different sets of independent measurements with N=1, 2, 4, 16, 32, 64 and 128 respectively. Data shown in Fig. 5 corresponds to the standard deviation at one particular and representative wavenumber, 1500 cm$^{-1}$. We repeated the same experience with two different spectral resolutions, 8 cm$^{-1}$ and 16 cm$^{-1}$ respectively. In both cases, a $1/\sqrt{N}$ dependence is observed, being characteristic of additive Gaussian noise. The inset in Fig. 5 shows a typical intensity spectrum. As it can be seen, the beam intensity is not constant across the measured spectral range. It is quite feeble at the extrema and shows a maximum in the central part, around 2500 cm$^{-1}$. The peak (at 2382.66 cm$^{-1}$) and the two bands (at 1545.06 cm$^{-1}$ and 3781.03 cm$^{-1}$) are due, respectively, to the carbon dioxide and water vapor present in the atmosphere. If the instrument was operated in a vacuum chamber, those absorption bands would disappear. Because the signal level is not constant, the noise level, $\sigma_0$, is higher at the ends than at the center of the measured spectral range. The error bars in Fig. 5 account for the uncertainty in the determination of $\sigma_0$. The dependence of the standard deviation with N, shows a clear advantage of averaging 16 to 32 independent measurements because the noise level can be considerably reduced to an acceptable value without requiring and excessive acquisition time. Increasing the number of accumulated measurements beyond 64, does not allow to improve signal quality significantly and makes measurements considerably long. In practice, the information from Fig. 5 is used to set the number N required to obtain a target value of $\sigma_0$. In our case we used to work with a target value lower than $1 \cdot 10^{-3}$.

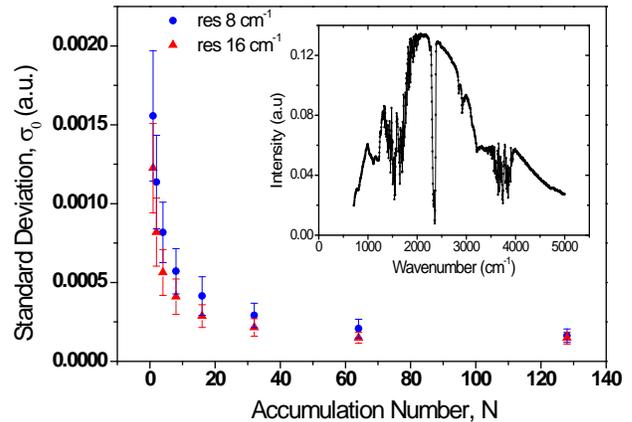

Fig. 5 Standard deviation as a function of the accumulation number, N. Spectral resolution was set to 8 cm$^{-1}$ (blue) and to 16 cm$^{-1}$ (red). Error bars represent the uncertainty in the determination of the standard deviation. The inset shows the intensity dependence of the intensity in a typical spectra with 8 cm$^{-1}$ resolution.

The time required to measure a full intensity matrix can vary from 4 to 8 minutes depending on the selected spectral resolution and the reflectivity of the sample. Once the value of $\sigma_0$ is known, it is just necessary to apply Eq.(5) to evaluate the uncertainty of each Mueller matrix element due to random noise. According to the results of Fig. 4, the minimum experimental uncertainty due to random noise propagation is estimated to be around $2 \cdot 10^{-3}$ (for Mueller matrix elements in the first and second rows), and about $5 \cdot 10^{-3}$ for the remaining matrix elements. The effect of systematic errors is not taken into account in our evaluation of $\sigma_0$ because it depends on a number of empirical parameters that may change from experience to experience or from day to day. There are many sources of systematic errors such as room temperature fluctuations that can affect the retardance created by the bi-prisms, the composition of the atmosphere (% of water vapour or $CO_2$ for instance) that may change its absorption, or drifts in the radiance of the infrared source among others. We have found that the main factor that influences the systematic errors in our experiences is the misalignment occurring when samples are mounted and removed from the sample holder during routine operation. To circumvent this problem a (He:Ne) laser diode is inserted in the optical path prior to each measurement to check that the sample is well positioned and that the beam follows the same path that was defined in the calibration step. As a result of a careful operation of the Mueller ellipsometer in a temperature stabilized room and in a controlled atmosphere, allow to bring the impact of the systematic errors to a level comparable to that of random noise previously discussed. Thus, we can estimate that the maximum average accuracy that can be reached with our experimental setup is about 0.005 for Mueller matrix elements in the first row and column; and around 0.01 for the remaining matrix elements. In the following we provide several examples.

### C. Results and discussion

#### 1.A Measurement of a linear polarizer

To evaluate the overall system performance we measured the Mueller matrix of a linear polarizer (a commercial grid polarizer) rotated at various orientations respect to the plane of incidence. The polarizer was placed after the PSG. Since the Mueller ellipsometer was set in the reflection mode, a gold mirror was placed on the sample-holder. The Mueller matrix of the polarizer was obtained multiplying the matrix of the system polarizer-mirror by the inverse of the matrix of the mirror measured without the polarizer. The results corresponding to a wavenumber of 1500 cm$^{-1}$ are shown in Fig. 6.

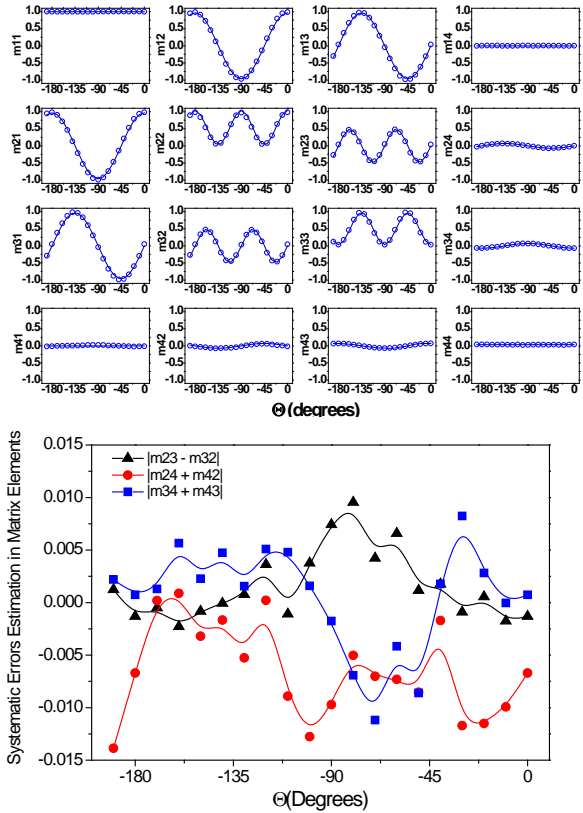

Fig. 6. Top panel Mueller matrix of a grid linear polarizer oriented at various azimuths. Dots correspond to experimental data at 1500 cm$^{-1}$ and solid line to the best-fitted model curve. Lower panel. Differences among some symmetric Mueller matrix elements indicated in the legend. Lines are just a guide to the eye.

Spectra was recorded with 16 cm$^{-1}$ resolution and 32 integrations. In those conditions, according to data in Fig. 5 and Fig. 6 the uncertainty due to random noise in Mueller matrix elements is expected to be approximately 0.0025. The absolute value of the rotation angle was known with an accuracy of about 0.3° for each measurement. To interpret the dependence of Mueller matrix elements with the polarizer azimuth, θ, we assumed that the polarizer behaved as an imperfect dichroic retarder with a small amount of depolarization. The imperfections caused a slight isotropic depolarization were represented by the parameter D. The model matrix could be written as:

$$P = \begin{pmatrix} 1 & -DI'_c C_\theta & -DI'_c S_\theta & 0 \\ -I'_c C_\theta & D(C_\theta^2 + I_c) & DC_\theta S_\theta (1-I_c) & -DI_s S_\theta \\ -I'_c S_\theta & DC_\theta S_\theta (1-I_c) & D(S_\theta^2 + I_c C_\theta^2) & DI_s C_\theta \\ 0 & +DI_s S_\theta & -DI_s C_\theta & DI_c \end{pmatrix}, \quad (6)$$

where $C_\theta = \cos(2\theta)$, $S_\theta = \sin(2\theta)$, $I'_c = \cos(2\psi)$, $I_c = \sin(2\psi)\cos(\Delta)$, and $I_s = \sin(2\psi)\sin(\Delta)$.

Apart from the azimuth and the depolarization, the matrix elements depend on the ellipsometric functions Ic', Is and Is which are, in turn, trigonometric functions of the ellipsometric angles Ψ and Δ. Experimental data was fitted to Eq. 6 model using as parameters Ψ, Δ and D giving as a result: Ψ=87.4±0.5°, Δ=294.2±0.5°, D=0.971±0.001. We found from the analysis of residuals that the average value of the differences among fitted and experimental data was 0.008. According to Eq. 6, there are some pairs of Mueller elements that have identical respective values. Therefore, calculating the differences among them is an heuristic way to evaluate the magnitude of systematic and non-systematic errors in experimental data. The lower panel of Fig. 6. shows the differences among three symmetric pairs of matrix elements. As can be seen in the lower panel of Fig. 6, there is a strong dependence of the values shown in there with the polarizer orientation. For azimuths comprised from -180° to 110° the values of the differences are constant and around 0.005, however, for azimuths comprised between -110° and -45°, the differences become higher, up to 0.01. In our opinion the parallelism between the faces of the polarizer was not perfect, thus causing beam wandering when the azimuth was modified. Beam wandering can create systematic errors because the beam is deviated from the path that initially followed during the calibration procedure. In our opinion for azimuths from -180° to -110° the magnitude of systematic errors is comparable or lower than that of random noise, but for azimuths between -110° and -45° systematic errors become dominant. Thus, in ideal conditions, a nominal accuracy of the instrument can be estimated to be about 0.005, which is the sum of two balanced contributions, noise bounded to 0.0025 and systematic errors bounded to about 0.0025. In non-ideal conditions, systematic errors can surpass random noise thus increasing the uncertainty of the measurements.

#### 2.Characterization of a thermally oxidized c-Si wafer

The second example consists of the measurement of a crystal silicon (c-Si) wafer covered with a thin film of amorphous silicon oxide (SiO$_2$) thermally grown on it, with a nominal thickness of 1050 nm. The Mueller matrix of this kind of samples (isotropic and non-depolarizing) can be expressed in terms of the ellipsometric angles Ψ and Δ as [27]:

$$M = \begin{pmatrix} 1 & -I'_C & 0 & 0 \\ -I'_C & 1 & 0 & 0 \\ 0 & 0 & I_C & I_S \\ 0 & 0 & -I_S & I_C \end{pmatrix} \quad (7)$$

As a representative example, Fig. 7 shows the experimental spectroscopic Mueller matrix at an angle of incidence of 68.2°. It can be seen that the measured matrix has the same structure as Eq. 7, with null block-diagonal elements.

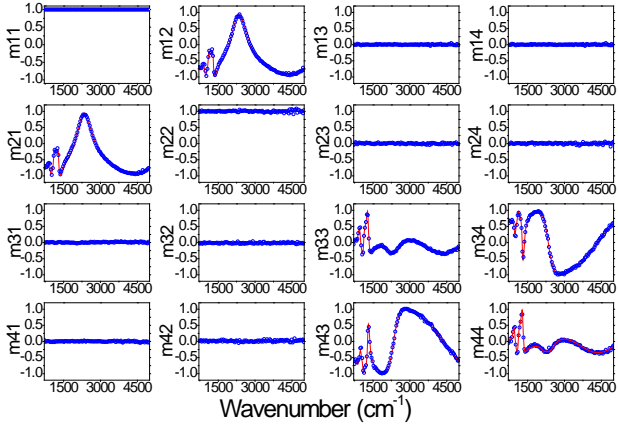

Fig. 7. Experimental (blue dots) and best-fitted (red line) spectroscopic Mueller matrices of a silicon wafer coated with a thermal $SiO_2$ film of nominal thickness 1050 nm.

A common practice to extract information from ellipsometric data, is to simulate the optical response of the sample expressed in terms of Mueller matrix elements or any other related observable, by a theoretical model. To do so, typical models allow to calculate the reflectivity of the sample in polarization parallel "p" and perpendicular "s" to the plane of incidence respectively, when it has been illuminated by beams polarized according to "p" and "s" directions respectively. Once these 4 reflectivity coefficients are calculated, it is straightforward to calculate the Jones and the Mueller matrix of the considered sample [27,28]. There are well stablished methods to calculate the optical response of samples consisting of a number of layers deposited on a substrate [29]. In general, theoretical models depend on a number of parameters characteristic of the sample, which must be adjusted to make the theoretical data to "fit" the measurements. A typical model for a stack of layers includes the angle of incidence, the thickness and complex refractive indexes of all the layers. The dispersion of the complex refractive index can be taken from reference tables or represented by a parameterized mathematical expression. Further discussion about data interpretation and modelling in the case of either, standard ellipsometry and Mueller ellipsometry, is given in specialized monographs [27,28,30]. Ellipsometric information taken at different angles of incidence may be useful when it comes to fit measurements with a model that depends on many parameters. As discussed in previous works [1,28,31], simultaneously fitting data taken at different angles of incidence allows to decrease linear correlations among the fitted parameters and thus, to improve the accuracy of the fitted values. We took measurements at different incident angles, (from 75° to 65° with a step of 2°). Experimental data was fitted with a simple mathematical model consisting of a semi-infinite c-Si substrate covered by a homogeneous $SiO_2$ layer. The dielectric functions of c-Si and $SiO_2$ were taken from a reference [32]. The fitted parameters were the $SiO_2$ layer thickness and angle of incidence. As previously mentioned, in conventional ellipsometry is common to use parameterized dielectric functions to represent the optical properties of the materials analyzed. We have preferred to work with reference tabulated values for two reasons: i) to reduce the number of parameters to be adjusted, and ii) because both, c-Si and $SiO_2$ are extensively studied materials, and their dielectric functions are well-known. Based on our experience, we knew that that the tabulated dielectric functions for c-Si and $SiO_2$ corresponded to those of the materials present in the sample. A detailed discussion of the parameterization of dielectric functions for c-Si and $SiO_2$ goes beyond the objectives of this paper, but this topic has been discussed in detail by Ossikovski *et al.* in [33]. The resulting best-fitted values for the $SiO_2$ layer thickness was 1046 ± 10 nm, in good agreement with the manufacturer specifications. The best fitted angles of incidence were 74.8°, 73.0°, 71.2°, 69.7°, 68.2° and 66.5°. Correlation between angle of incidence and thickness was low, about 0.3. To illustrate the dependence of the polarimetric data with the angle of incidence we plotted in Fig. 8 the experimental and best-fitted values. Because the block diagonal elements are null, and some of them are redundant, only the elements [1,2], [3,4] and [4,4] are shown. An excellent agreement is obtained between experiment and theoretical values, providing the usefulness of the model and the accuracy of data measured with the ellipsometer. In the plots two types of spectral features can be seen. The first one, around 2500 cm$^{-1}$, is related to a Bragg type interference between beams partially reflected by the first and second faces of the $SiO_2$ layer. The spectral position of this feature is mainly determined by the $SiO_2$ layer thickness, but changing the angle of incidence makes the position of the interference to slightly shift by 75 cm$^{-1}$. Changing the angle of incidence modifies the path length of the beam inside the layer and thus the wavenumber for which the Bragg condition is fulfilled.

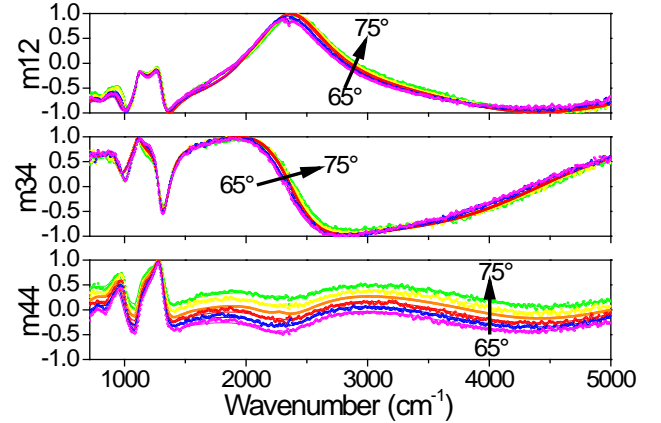

Fig. 8. Spectroscopic Mueller elements [1,2], [3,4] and [4,4] as a function of the angle of incidence. Arrows indicate the dependence of spectroscopic values with the angle of incidence.

The second spectral feature, between at 1000 and 1300 cm$^{-1}$, is related to the absorption due to the different stretching modes of the Si-O bond. In particular, the signatures of a transversal optical mode at 1130 cm$^{-1}$ and a longitudinal optical mode at 1260 cm$^{-1}$ are well visible [33]. The spectral position of this band, contrarily to that of the Bragg interference, does not depend on the angle of incidence.

### D. Conclusions

We have built a new spectroscopic infrared Mueller ellipsometer based on achromatic optical elements. The optical set-up consists of a polarization states generator, a PSG, and a polarization states analyzer, a PSA, both being identical. The optical design is based on an objective criterion, the maximization of the condition number of the matrices representing the PSG and the PSA, which allows minimizing the propagation of random and systematic errors to the measurements. The eigenvalue method has been applied to calibrate the PSG and the PSA without ambiguities because it does not need to model the optical elements forming the Mueller ellipsometer. We have included in this article several examples to illustrate the quality of the polarimetric measurements, compatible with that of commercial ellipsometers.

Acknowledgements

The authors wish to express their acknowledgment to Dr. Y. Bonnassieux, and Dr. P. Bradu for their help in finding funds and providing part of the components. Moreover, the authors wish to express their gratitude to Mr. D. Clément, Mr. C. Jadaud and Mr. F. Farci for their help in designing and building the non-standard opto-mechanical pieces used. The development of the broadband Mueller ellipsometer was a project included in the Master's Program at Ecole Polytechnique. The authors, wish to thank Mr. R. Landon, supervisor of one the programs, for his help in communication and integration of students in the project.